\documentclass[journal]{IEEEtran}

\usepackage{amsmath}
\usepackage{eulervm}
\usepackage{graphicx}
\usepackage{caption}
\usepackage{url}
\usepackage{multirow}
\graphicspath{{Figures/}}
\usepackage{subcaption}
\usepackage[sort,compress]{cite}
\usepackage{times}
\usepackage{algorithm}
\usepackage{verbatim}
\usepackage{sidecap}
\usepackage{array}
\usepackage{float}
\usepackage{comment}
\usepackage{minibox}
\usepackage{color,soul}
\usepackage{enumerate}

\usepackage{xcolor,soul}
\colorlet{shadecolor}{yellow}
\usepackage{mathtools}
\usepackage{multicol}
\usepackage{dblfloatfix}
\input epsf

\def\foo nife{Ni\textsubscript{80}Fe\textsubscript{20}} 
\def\bst bst{(Bi\textsubscript{0.2}Sb\textsubscript{0.8})\textsubscript{2}Te\textsubscript{3}}
\def\cofeb cofeb{Co\textsubscript{40}Fe\textsubscript{40}B\textsubscript{20}} 
\def\bse bse{BiSe\textsubscript{2}}

\author{Arifa~Hoque~and~Sanjukta~Bhanja\\
\thanks{Authors are with the Department
of Electrical Engineering, University of South Florida, Tampa,
FL, 33612, USA, e-mail: arifahoque@usf.edu}}%

\begin{document}

\title{Controllable Domain Wall Memories with Magnetic Topological Insulator}

\maketitle

\begin{abstract}

\textbf{Domain wall memories have undergone several changes over the years for faster shift, read, and write operations; however, fundamental issues persist due to creating pinning sites topographically along the nanowire. The deformity in notches creates non-uniform pinning strength, leading to multiple faults during shift operation. This study proposes a novel approach to address these challenges and gain greater control over domain manipulation for advanced applications in Boolean and non-Boolean paradigms. Utilizing the magnetic topological insulator (MTI) can create pinning sites without potentially faulty topographical notches made through complex lithography. Moreover, applying an external current enables precise control over the pinning potentials at these sites. Micromagnetic simulations validate the effectiveness of MTI-based pinning sites, showcasing their potential for future applications. Our approach introduces an alternative method for creating pinning sites by cross-architecture of ferromagnetic nanowires and MTI nanobars, inducing exchange interaction at their intersection points. This method offers simplicity of fabrication and enables control over the pinning strength by external current.} 

\end{abstract}

\section{Introduction}
\label{sec:intro} 

\IEEEPARstart{D}{}omain Wall Memories (DWMs) are a class of spintronic memories often tagged as ``Racetrack Memories''~\cite{parkin2008magnetic}, which store multiple bits of data laterally in long ferromagnetic nanowires. Domain wall memories possess many positive attributes, including SRAM-class access speed, reduced static power due to non-volatility, high endurance, CMOS compatibility, and higher data retention. A DWM nanowire holds each bit of data by magnetic domains separated by domain walls, which are gradual reorientations of magnetic moments. The efficacy of domain wall memory operations depends on the propagation of domain walls, which is stochastic in nature~\cite{akerman2010stochastic}. To ensure reliable memory operations, it is essential to precisely control the motion of domain walls by pinning them at equally spaced locations. Uniform pinning of domain walls is critical for reliable operation. Without pinning sites, domain walls can move rapidly and uncontrollably, leading to unpredictable behavior and hindering device functionality. Structural modifications, such as local constrictions or protrusions, are commonly used to achieve stable DW positions in strips. Traditional pinning sites often require complicated lithography processes, such as the creation of notches and zigzag patterns on ferromagnetic nanowires~\cite{huang2009domain, zhou2017current}. This complexity adds to manufacturing overhead and may limit scalability. Additionally, pinning site shape and size variations can lead to non-uniform pinning effects, affecting device performance and reliability~\cite{roxy2022pinning}. A comparatively higher pinning strength is a requirement for better data endurance, but it leads to increased power consumption. On the other hand, low-power memory operations need shrunk nanowires with weaker pinning. A balance between both is the research question of this work.

Fabricating these structural pinning sites uniformly at the nanoscale is challenging and irreversible. Once the pinning sites are formed, their positions cannot be changed, limiting the flexibility and adaptability of DW-based devices. This limitation is particularly problematic for neuromorphic applications, where programmable multilevel states are essential to emulate synaptic functions. Thus, there is a critical need for position-reconfigurable pinning methods to realize DW motion-based neuromorphic applications. Dynamic adjustment of pinning strength offers a potential solution to the challenges associated with controlling DW motion. Changing the DW position as needed becomes feasible by temporarily reducing the pinning strength. Various solutions have been proposed to address the challenges associated with traditional pinning methods in magnetic domain wall devices. These include manipulating exchange bias~\cite{polenciuc2014domain} to influence pinning strength, modifying ferromagnetic materials using techniques such as ion irradiation or doping~\cite{king2014local,burn2014control,rantschler2007effect,faulkner2007rapid,jin2017tuning}, and using electrical gating~\cite{breitkreutz2014controlled,franken2013voltage,bernand2013electric} to precisely pin domain walls within the device.

In recent works, introducing pinning sites with dynamically adjustable strength has been proposed to enable precise control over DW motion. One method harnesses stray magnetic fields from neighboring ferromagnetic elements to generate reconfigurable pinning effects. By controlling the magnetic state of neighbors, the strength of the resulting pinning sites can be modified~\cite{metaxas2009periodic,o2011tunable,hiramatsu2013domain,metaxas2013spatially,franken2014beam,van2014control}. Hurst \textit{et al.} demonstrated that highly localized stray magnetic fields around the core of magnetic vortices enable toggling between strong and weak DW pinning strengths~\cite{hurst2017reconfigurable}. This approach offers tunability without the need for precise fabrication of nanostructures. Another recent approach~\cite{lee2023position} is to leverage the dipolar interaction between domain walls in magnetic double layers to create a pinning potential barrier. The domain wall motion in the top nanowire generates a dipolar field that repels the walls in the bottom layer, effectively pinning them. Notably, the position of the pinning site can be modulated by manipulating the walls, enabling reconfigurable pinning for DW motion.

Topological insulators are a unique state of matter that have the potential to be used in next-generation spintronic devices. This work explores the interaction between the accumulated surface state electrons in magnetic topological insulators (MTIs) and the magnetization of FM nanowires, which introduces advanced spin-moment manipulation and can lead to more efficient domain wall memory devices. The proposed device uses current flow through the surface states of MTIs to create a controllable pinning potential that can be manipulated by external energies. Also, the reconfigurable pinning strength allows quantized pinning sites along the nanowire to be used in near-analog computation. The paper discusses the fundamentals of MTIs, the physics of pinning and de-pinning domain walls, and the structure and working principle of the proposed device. The experiments demonstrated the controllability of pinning potentials, which could have potential use as non-Boolean computing elements.

 \section{Magnetic Topological Insulator}
Topological Insulators (TIs) exhibit distinctive quantum-mechanical characteristics primarily stemming from their atypical surface states. Despite the insulating nature of a TI's bulk, its surface behaves like a conductor due to band inversion at the interface~\cite{liu2010model,zhang2009topological}. The origin of this band inversion lies in the presence of strong spin-orbit coupling (SOC) within TIs. By introducing magnetism to TIs through methods such as impurity doping with magnetic elements, proximity coupling to magnetic insulators (MIs), or adding magnetic layers, time-reversal symmetry (TRS) can be broken. This TRS breaking induces a bandgap in the surface states, resulting in the Quantum Anomalous Hall Effect (QAHE), characterized by one-dimensional chiral edge conduction. Inside the bulk of a 3D MTI, SOC has minimal impact due to the crystal's symmetry~\cite{zhang2009topological}, and experimental observations have confirmed the presence of a 0.3 eV bandgap in the bulk region~\cite{mellnik2014spin}. Nonetheless, the 3D MTI's surface exhibits pronounced SOC effects, pushing the conduction band downward and raising the valence band, causing these bands to meet at the surface and resulting in spin-polarized surface states.

\subsection{Electrical Transport and Torque Generation}

When an in-plane current traverses the conductive surface states of a topological insulator, the forward-going electron states outnumber the backward-going states due to the helical locking of the spin-momentum orientations in the surface state. This implies that a non-equilibrium surface spin accumulation accompanies the charge flow. If the TI surface state spins ($\Vec{S}$) are coupled with the magnetic moments ($\Vec{m}$) of an adjacent ferromagnetic layer through $H_{ex} = -\Delta_{ex} \Vec{m} \cdot \Vec{S}$, where $\Delta_{ex}$ is the exchange coefficient, the spin polarization on the TI surface leads to a spin-transfer torque on the magnet. This torque is related to the Rashba–Edelstein effect~\cite{edelstein1990spin}, which facilitates the helical locking of spin and momentum at the TI surface, resulting in a high charge-to-spin current conversion ratio~\cite{mellnik2014spin,liu2010model}.

Two types of torque act on the FM layer due to spin transfer: the field-like torque and the Slonczewski-like torque~\cite{fan2014magnetization,fischer2016spin}. The field-like torque is proportional to the exchange coupling~\cite{yokoyama2010theoretical} and depends on the non-equilibrium spin density ($\vec{S_d}$)—in other words, on the spin polarization of the TI surface—and is of the form $\vec{T}_{FL} = \Delta_{ex}(\vec{m} \times \vec{S_d})$. The non-equilibrium spin density characterizes the diffused spin current density in the ferromagnetic layer as $J_s = -D \vec{\nabla} \cdot \vec{S_d}$, where $D$ is the diffusion coefficient of spin inside the ferromagnetic layer. The spatial change of spin current due to diffusion leads to the generation of Slonczewski-like torque, which can be expressed~\cite{manchon2012spin} by Eq.~\ref{eqn:STT torque-1}, where the spin current diffuses in the $z$-direction into the adjacent ferromagnetic layer.

\begin{equation}
    \Vec{T}_{STT} = \frac{1}{d} \int_0^d \left( -\vec{\nabla}_z J_s - \frac{1}{t_{sr}} \vec{S_d} \right) dz
    \label{eqn:STT torque-1}
\end{equation}

\noindent $d$ and $t_{sr}$ in Eq.~\ref{eqn:STT torque-1} refer to the thickness of the ferromagnetic layer and the spin-relaxation time, respectively. Under steady-state conditions, the spin diffusion ($\vec{S_d}$) into the domain wall nanowire from the MTI nanobar is expressed by the equation~\cite{kurebayashi2014antidamping}:

\begin{equation}
    \Vec{\nabla} J_s = -\frac{1}{t_j} \vec{S_d} \times \Vec{m} - \frac{1}{t_{\phi}} \vec{m} \times (\vec{S_d} \times \vec{m}) - \frac{1}{t_{sf}} \vec{S_d}
    \label{eqn:STT torque-2}
\end{equation}

\noindent where $t_j$, $t_\phi$, and $t_{sf}$ refer to the spin precession, decoherence, and relaxation times, respectively. If the thickness of the FM layer exceeds the spin diffusion length, the boundary conditions become $J_s(0) = pJ_{s0}$ (with spin-injection efficiency $p$) and $J_s(d) = 0$. The spin diffusion density ($S_d$) can be calculated from Eq.~\ref{eqn:STT torque-2} by applying these boundary conditions under the assumption of spin variation in the z-direction, and can be simplified as:

\begin{equation}
    S_d = \frac{pJ_{s0}L}{D} \left( \frac{\sinh\left( \frac{d-z}{L} \right)}{\sinh\left( \frac{d}{L} \right)} \right)
\end{equation}

\noindent $L$ in the above equation is a function of the spin relaxation length ($\lambda_{sf}$), spin precession length ($\lambda_j$), and spin decoherence length ($\lambda_\phi$), and is defined as $\frac{1}{L^2} = \frac{1}{\lambda^2_{sf}} + \frac{1}{\lambda^2_{\phi}} - \frac{i}{\lambda^2_{j}}$. The Einstein-Smoluchowski relation connects the spin diffusion coefficient ($D$) with the spin relaxation length~\cite{manchon2012spin} as $Dt_{sr} = \lambda^2_{sf}$. Combining all definitions into Eq.~\ref{eqn:STT torque-1} and simplifying yields the torque expression:

\begin{equation}
    T_{STT,\parallel} + iT_{STT,\perp} = \frac{pJ_{s0}L^2}{d} \left( \frac{1}{\lambda^2_{\phi}} - \frac{i}{\lambda^2_j} \right) \frac{\cosh\left( \frac{d}{L} \right) - 1}{\sinh\left( \frac{d}{L} \right)}
    \label{eqn:T_STT}
\end{equation}

With the FM layer thickness larger than the spin-diffusion length, this Slonczewski-like torque becomes several orders of magnitude higher than the field-like torque~\cite{keller2012facing}. In this work, we harness the exchange interaction between the TI surface states and the local magnetic moments of the FM layer to pin the domain wall at the intersection. The torques on the FM layer from the TI surface can control the de-pinning of the walls, hence providing robust reconfigurability of the domain wall motion.

\section{Domain Wall Memory with MTI}
This section describes the proposed device architecture, current-induced depinning, and standard memory operations.

\subsection{Proposed Device Structure}

\begin{figure}[!t]
\center
\includegraphics[width=3.4in]{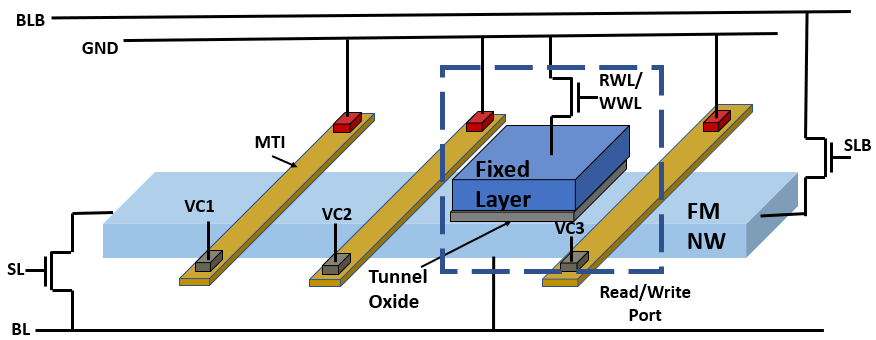}
\caption{Structure of the proposed device (not drawn to scale). MTI nanobars lie orthogonal on top of the nanowire to create pinning sites at the cross-section. The dashed box area represents the read/write port of the nanowire.}
\label{fig:structure}
\end{figure}

\noindent Fig.~\ref{fig:structure} illustrates the conceptual structure of the proposed device with four domains, similar to conventional DWM nanowires, except MTI nanobars replace the notches. The fundamental cell contains a planar ferromagnetic nanowire (in the x-direction), which holds the data. On top of the FM nanowire, a set of equally spaced MTI nanobars are placed to create pinning sites at the cross-sections. Each MTI nanobar contains ports to provide external excitations (voltage/current) along the nanobars to control pinning potentials. One of these ports is connected to ground, while the others (VC1-VC3) provide external voltage/current. Similar to conventional DWM cells, there are two shift ports available at the extremities of the FM nanowire to enable bi-directional shifting of data. The data-shift operation is controlled by two access transistors with control signals labeled SL and SLB. Two complementary bit-lines, BL and BLB, provide necessary excitations for regular memory operations, i.e., read, write, and shift of data. Traditionally, the read and write operations of domain wall nanowires are similar to spin-transfer torque magnetic random access memory (STT-MRAM) and are done by forming magnetic tunnel junction (MTJ) ports where the free layer is a domain of a nanowire and a thin oxide layer is sandwiched between the free layer and a ferromagnetic fixed layer. Such access ports can be shared between read and write operations; however, dedicated read and write ports are also common~\cite{zhao2012magnetic,foerster2014domain}. An access transistor is integrated with an MTJ port to control the read/write operation. In Fig.~\ref{fig:structure}, a shared access port is demonstrated inside the dotted box. The RWL/WWL control signal enables the desired operation. Over the last few years, significant research effort~\cite{venkatesan2013dwm,2013sunDac} has been made to find low-power write operations. Since this work targets controllable domain wall motions, we limit our focus to the shift operation.

\begin{figure}[H]
    \centering
    \includegraphics[width=2.6in]{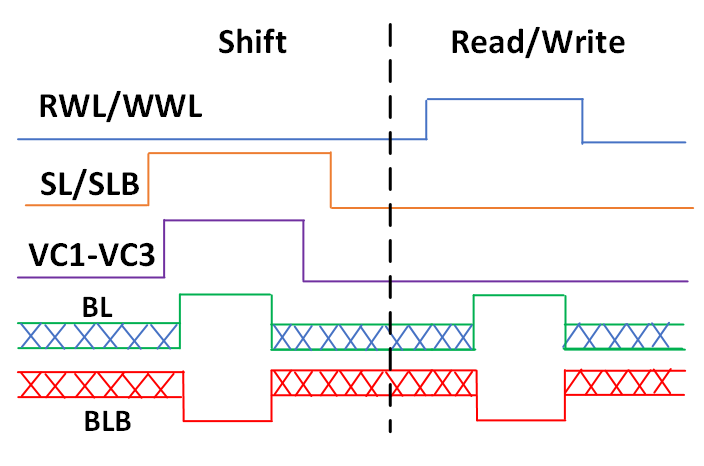}
    \caption{Control signals for read, write, and shift operations.}
    \label{fig:signals}
\end{figure}

\subsection{Current-induced Domain Wall Motion}

\noindent During current flow through the FM nanowire, the 4s conduction band electrons interact with the 3d band electrons of magnetic domains, generating an exchange interaction torque~\cite{berger1984exchange}. At a domain wall, the magnetization flips over a plane parallel to the wall. As a result, the conduction electrons bend at the wall while crossing it, amounting to a transfer of momentum between the passing electrons and the local magnetic moment of the wall. Due to energy conservation, the electrons exert a force on the wall, which can move it with sufficient current density~\cite{tatara2004theory}. The micromagnetics of shifting a DW is governed by the Landau-Lifshitz-Gilbert equation with the inclusion of current-induced torques~\cite{tatara2008microscopic} (assuming the current flows into the x-direction), as:

\begin{equation}
\frac{d\vec{M}}{dt} = -\gamma\vec{M} \times \vec{H}_{eff} + \alpha\vec{M} \times \frac{d\vec{M}}{dt} - v_j \frac{\partial\vec{M}}{\partial x} + \beta v_j \vec{M} \times \frac{\partial\vec{M}}{\partial x}
\label{eq:LLG}
\end{equation}

\noindent where $\vec{M}$, $\vec{H}_{eff}$, $\alpha$, $\gamma$, and $\beta$ are the magnetization orientation, effective field, Gilbert damping constant, gyromagnetic ratio, and non-adiabatic spin-torque coefficient, respectively. The last two terms that are functions of $\partial\vec{M}/ \partial x$ represent the current-induced torques responsible for DW shifting, with a shifting velocity of $v_j$.

\subsection{Standard Memory Operations}

\noindent Like traditional domain wall memories, the proposed device can shift stored data bi-directionally. Fig.~\ref{fig:signals} plots the states of the control signals for shift and read/write operations. During data shifting, SL/SLB are set to VDD to establish the current path. Consequently, an external current is applied to control ports of MTI nanobars, VC1-VC3, to lower the pinning potential. At the same time, the bit-lines, BL and BLB, are raised to VDD and pulled down to GND, respectively, from high-impedance (Z) states for the left-shift and vice versa for the right-shift of the data. During the entire operation, the control signal RWL/WWL is set to GND to separate the read/write paths from the shift path. After a successful shift operation, SL, SLB, and VC1-VC3 are pulled down to GND to reset the pinning potential. Similarly, during the read or write operation, SL, SLB, and VC1-VC3 are kept at GND, and RWL/WWL is raised to establish a stable read/write path. The bit-line BL is set to VDD to provide a write current for writing or a smaller read current for reading. The `1' and `0' states have two different resistance levels depending on the direction of magnetic moments of the free and fixed layers, and when a read current is applied, a voltage drop is observed across the read port. Data can be read by measuring this voltage drop. During the write operation, the applied current is high enough to alter the magnetization using spin-transfer torque (STT). We refer readers to~\cite{kumar2018domain} for a more detailed explanation of read, write, and shift operations of domain wall memories.

\section{Experimental Setup}
\label{sec:setup}

\begin{figure}[!t]
    \centering
    \begin{subfigure}[b]{0.48\textwidth}
        \includegraphics[width=0.98\linewidth]{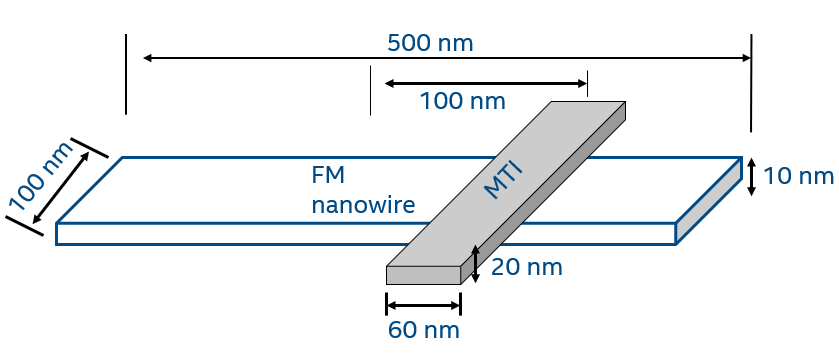}
        \caption{Device geometry used for simulation.}
        \label{fig:exp_setup_device}
    \end{subfigure}
    \begin{subfigure}[b]{0.24\textwidth}
        \includegraphics[width=0.98\linewidth]{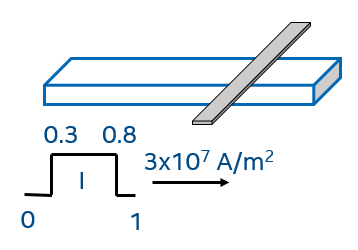}
        \caption{Setup for pinning experiment}
        \label{fig:exp_setup_pin}
    \end{subfigure}
    \begin{subfigure}[b]{0.24\textwidth}
        \includegraphics[width=0.98\linewidth]{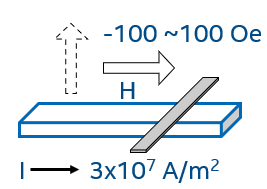}
        \caption{Setup for AMR calculation}
        \label{fig:exp_setup_amr}
    \end{subfigure}
    \begin{subfigure}[b]{0.48\textwidth}
        \includegraphics[width=0.98\linewidth]{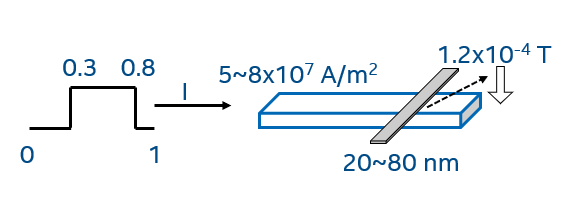}
        \caption{Depinning and critical current calculation.}
        \label{fig:exp_setup_depin}
    \end{subfigure}
    \caption{Experimental setup for pinning and depinning domain walls. All time stamp is in nanoseconds(ns). All distances are in nanometers(nm). Pictures not drawin to scales.}
    \label{fig:exp_setup}
\end{figure}

A finite-element micromagnetic model has been developed to simulate the behavior of a domain wall memory cell that utilizes magnetic topological insulator (MTI) nanobars as pinning sites. The model represents the ferromagnetic nanowire with periodically localized modifications in its magnetic properties to emulate the interface between the nanowire and the MTI nanobars. During domain wall shift operations, the torque induced at the FM-MTI interface is captured through a localized effective field, as detailed in Section IV. The micromagnetic dynamics of the system are governed by the Landau-Lifshitz-Gilbert (LLG) equation, presented in Eq.~\ref{eq:LLG}. We performed micromagnetic simulations using a GPU-enabled finite-difference solver, Mumax3~\cite{mumax} which performs the time integral simulations of the LLG equation. The effective field, $H_{eff}$, in Eq.~\ref{eq:LLG} includes the exchange, demagnetization, anistropy, Zeeman and the interfacial exchange coupling energy at the FM-MTI interface. We used a fixed time step of 10 femtoseconds (fs) to ensure numerical stability during domain wall dynamics. The total simulation time varies in different experiments and is listed in section~\ref{sec:res}. For energy minimization, we allowed the system to relax under zero external field until the total energy change was below 1×10\textsuperscript{-7}~J/m\textsuperscript{3} per iteration, ensuring a stable magnetization configuration. Fig.~\ref{fig:exp_setup} portraits the simulation setup for all the experiments done in this work.

\begin{table}[H]
\centering
\caption{Material properties of the nanowire and the NW-MTI interface used in simulation~\cite{reza2019modeling}.}
\begin{tabular}{|l|ll|}
\hline
Properties                                         & \multicolumn{2}{l|}{Value}                                                         \\ \hline
Material                                           & \multicolumn{1}{l|}{Nanowire}             & NW-MTI interface                \\ \hline
Sat. Magnetization (Ms) (A/m)                & \multicolumn{1}{l|}{$0.8\times10^6$} & $0.86\times10^6$        \\ \hline
Exchange Coeff. (A\_\{ex\}, J/m)              & \multicolumn{1}{l|}{$2\times10^{-11}$  } & $1.3\times10^{-11}$        \\ \hline
Easy Axis                                          & \multicolumn{1}{l|}{x-axis}                    & 45\textasciicircum{}0 with x-axis \\ \hline
Anisotropy Coeff. (Ku, J/m\textasciicircum{}3) & \multicolumn{1}{l|}{500}                       & 10\textasciicircum{}4             \\ \hline
Damping Constant, alpha                            & \multicolumn{1}{l|}{0.02}                      & 5.4*10\textasciicircum{}-4        \\ \hline
In-plane Spin Hall Angle                           & \multicolumn{1}{l|}{N/A}                       & 1.1                               \\ \hline
Out-of-plane Spin Hall Angle                       & \multicolumn{1}{l|}{N/A}                       & 1.03                              \\ \hline
\end{tabular}
\label{tab:parameters}
\end{table}

\noindent The first experiment intends to demonstrate the existence of the pinning of a doamin wall  using a magnetic topological insulator through micromagnetic simulation. We experimented with ferromagnetic nanowires exhibiting both in-plane and out-of-plane anisotropy. We consider ferromagnetic nanowires with in-plane magnetic anisotropy (IMA), made of \foo nife, and MTI nanobars, made of Co-doped \bse bse~\cite{zhang2014electrical,kumar2025co,reza2019modeling}. To demonstrate the similar pinning in nanowires with perpendicular magnetic anisotropy (PMA), we consider \cofeb cofeb as the nanowire material, and \bst bst~\cite{liu2023magnetic,reza2019modeling} as MTI nanobars. However, the following experiments were only carried out with IMA nanowires. The pinning and depinning physics in both IMA and PMA nanowires with MTI nanobars is similar; thus, in this work, we present the results for the IMA case, while our future work will include PMA nanowires. The material properties of the FM nanowire and the FM-MTI interface used for the simulation are listed in Table.~\ref{tab:parameters}. We used a cell size of 2$\times$2$\times$2 nm\textsuperscript{3} in our simulation, which is smaller than the thermal exchange length of 5.3 nm of \foo nife~\cite{abo2013definition} to ensure that the thermal fluctuation is negligible. 

Fig.~\ref{fig:exp_setup_device} shows the device setup for experiments in this work. The domain wall nanowire is 500 nm long, 100 nm wide, and 10 nm thick. One MTI nanobar of 60 nm wide and 20 nm thick is placed on top of the nanowire to pin the domain wall at the interface. The center-to-center distance in x-direction between the nanowire and the nanobar is 100 nm. To realize the domain wall pinning at the FM-MTI interface, an external current of 3x10\textsuperscript{7} A/cm\textsuperscript{2} is applied in the x-direction through the nanowire. During this experiment, there is no current flow through the MTI nanobar. The current pulse width is 1 ns with 50\% duty cycle as shown in Fig.~\ref{fig:exp_setup_pin}.

The second experiment is designed to realize the domain wall pinning at the interface by measuring the anisotropic magneto-resistance (AMR) of the nanowire with and without the MTI nanobar. The dimension of nanowire and the nanobar is similar to the first experiment. Fig.~\ref{fig:exp_setup_amr} shows the setup for this experiment where an external current is applied in the x-direction, and an external field ranging from -100 Oe to 100 Oe is applied parallel and perpendicular to the current flow. 

 The next experiment demonstrates the depinning of the domain wall by the torque exerted by the MTI nanobar into the nanowire at the interface. The amount of torque generated due to spin injection depends on the current flowing through the MTI nanobar and is calculated using Eq.~\ref{eqn:T_STT}. It is worth noting that, since the MTI is an electrical insulator and the nanowire is a conductor, there exists a leakage current from the MTI nanobar into the nanowire during current flow through the nanobars. Typically, such an arrangement with regular insulators results in significant leakage; however, due to strong spin–orbit coupling (SOC) and the resulting band inversion, the surface of the MTI layer becomes highly conductive, which significantly reduces the leakage through the nanowire.

Due to the proximity between ferromagnetic nanowires and topological insulators, time-reversal symmetry in the MTI is broken, leading to the Quantum Anomalous Hall Effect (QAHE). This effect creates gapless boundary states on the surface that carry dissipationless currents~\cite{yan2024rules}, leading to very small current shunting through the nanowire. To further reduce the leakage current, we increase the thickness of the MTI layer to 20~nm to decrease its sheet resistance and use a 10~nm thick \foo nife layer to increase the sheet resistance of the nanowire. The thickness of \foo nife can be further reduced to achieve a higher resistance path along the nanowire. Based on the calculations in~\cite{reza2019modeling}, we incorporate 10\% leakage current from the MTI to the FM layer in our experiments. The current values reported in this work represent the actual currents applied through the nanobars, while we reduce them by 10\% in the simulation to account for leakage current.

Fig.~\ref{fig:exp_setup_depin} shows the setup to demonstrate for this experiment showing the depinning and calculating the critical current for shifting. The torque is modeled as an external field and included in the $H_{\mathrm{eff}}$ term in Eq.~\ref{eq:LLG}. The details of modeling this torque as a field are explained in Section~\ref{sec:res_depin}. In this experiment, we also calculate the critical current required to depin and move the domain wall away from the interface in the absence of the exerted torque, for a given width of the nanobar. This implies that the pinning strength depends on the width of the MTI nanobar. These results are also important for manipulating domain wall motion in computing applications. Here, we varied the width of the nanobar from 20 to 80~nm in 10~nm steps, while keeping the FM geometry constant, similar to the first experiment. The surface charge was removed from the left and right boundaries of the nanowire to mimic an infinitely long nanowire, allowing the domain wall to travel beyond the physical boundaries.

\section{Results and Discussion}
\label{sec:res}

\subsection{Domain Wall Pinning by MTI}

\begin{figure}[H]
\centering
\includegraphics[width=3 in]{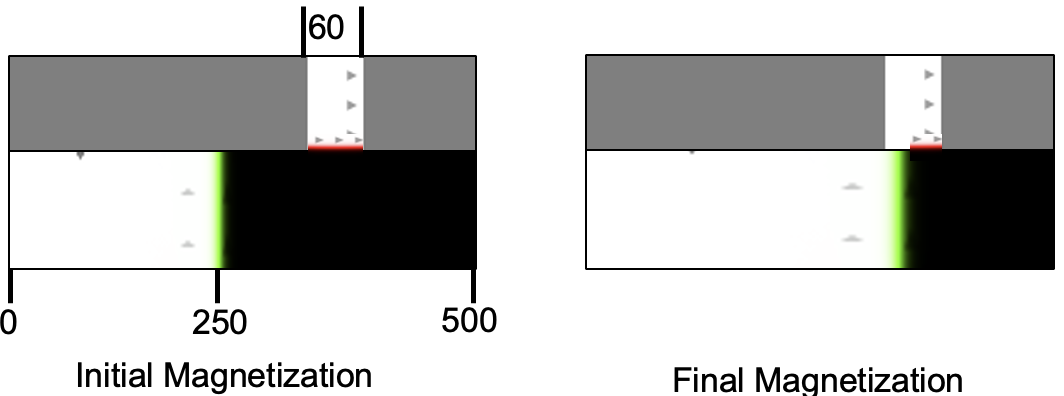}
\caption{Domain walls pinned at PMA FM-MTI crosspoints. (Left) Initial Magnetization; (Right) Final Magnetization.}
\label{fig:PMA-pinning}
\end{figure}

\begin{figure}[H]
     \centering
     \begin{subfigure}[b]{0.24\textwidth}
         \centering
         \includegraphics[width=0.98\textwidth]{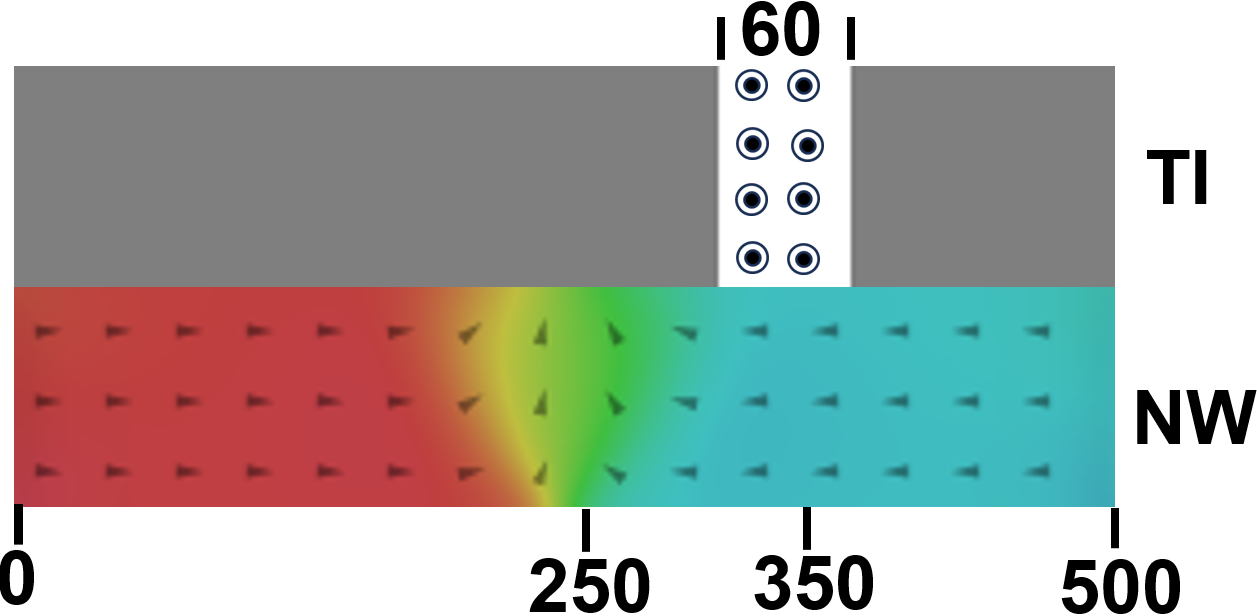}
         \caption{Initial Magnetization (T=0)}
         \label{fig:Tilt-initial}
     \end{subfigure}
     \hfill
     \begin{subfigure}[b]{0.24\textwidth}
         \centering
         \includegraphics[width=0.98\textwidth]{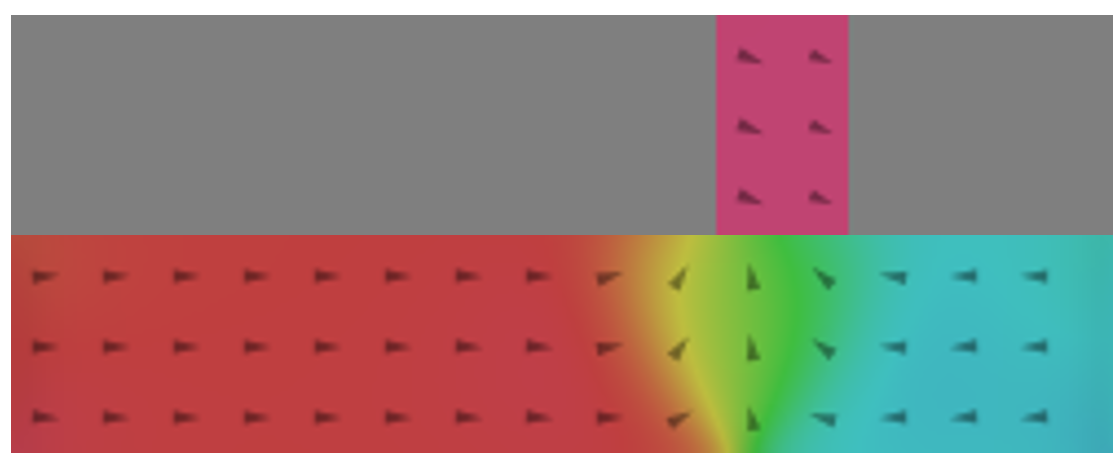}
         \caption{Final Magnetization (T=1 ns)}
         \label{fig:ATilt-final}
     \end{subfigure}
        \caption{Domain wall pinning at the MTI-FM interface. All values are in nm. The gray area of the TI represents a vacuum.}
        \label{fig:Tilt}
\end{figure}

\noindent At the cross-points of the proposed device, the direct exchange interaction between the MTI and FM nanowire generates a pinning potential to hold the domain wall (DW). Fig.~\ref{fig:PMA-pinning} and Fig.~\ref{fig:Tilt} depict the domain wall pinning at the FM-MTI interface in both PMA and IMA nanowires, respectively. 

At the beginning of the experiment (time, T = 0), the FM nanowire is initialized with a domain wall at the center, as shown in Fig.\ref{fig:Tilt-initial}, and is allowed to relax for 0.3 ns. The initial magnetization of the MTI layer is set to the (0,0,1) direction. After relaxation, at T = 0.3 ns, a shift current of 3x10\textsuperscript{7} A/cm\textsuperscript{2} is applied in the x-direction for 0.5 ns. Subsequently, the domain wall begins to move and eventually becomes pinned at the center of the FM–MTI interface due to the exchange interaction at the intersection. Fig.~\ref{fig:ATilt-final} shows the final magnetization of the nanowire at T = 1 ns, after allowing 0.2 ns for magnetization relaxation. The red, blue, and green areas in Fig.\ref{fig:Tilt} represent +x, –x-directed magnetization, and the +y-directed domain wall, respectively. Fig.~\ref{fig:PMA-pinning} shows the initial and final magnetization of the nanowire with PMA in a similar experiment. The white and black regions correspond to +z and –z-directed magnetization, while the pink and green regions indicate the domain walls.

Jin~\textit{et al.}\cite{jin2019tilted} experimentally demonstrated how tilted magnetization at pinning sites can effectively pin domain walls with comparable strength. Similarly, Polenciuc \textit{et al.}~\cite{polenciuc2014domain} reported pinning sites created by exchange bias between ferromagnetic and antiferromagnetic (AF) layers. However, both configurations lack programmability in pinning strength and require traditional energy-intensive domain wall depinning techniques, such as applying a magnetic field or passing current through the shift port. Our proposed device leverages the same principle for domain wall pinning but enables programmable pinning strength and STT-assisted depinning.

\begin{figure}[]
     \centering
     \begin{subfigure}[b]{0.35\textwidth}
         \centering
         \includegraphics[width=0.95\textwidth]{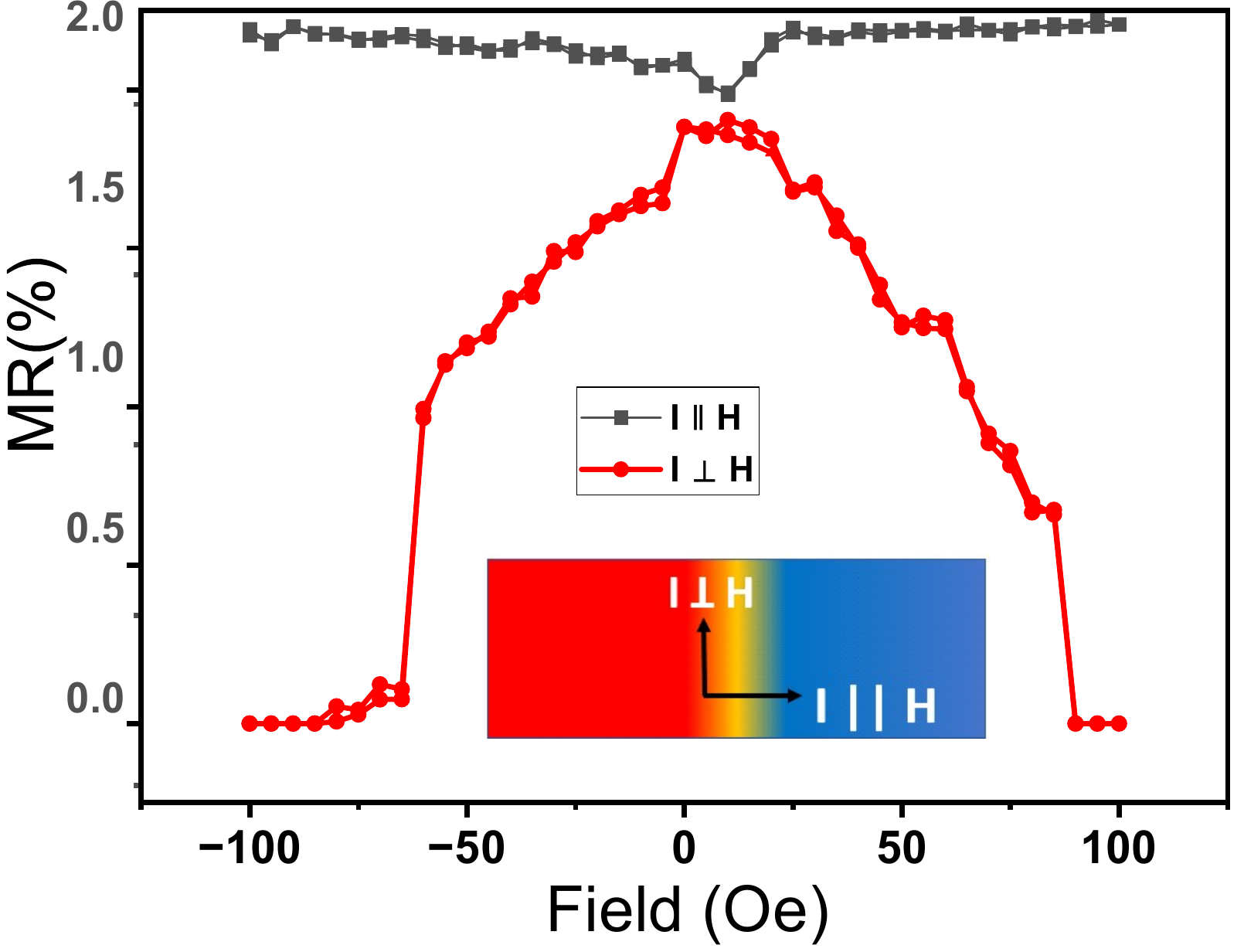}
         \caption{AMR without TI}
         \label{fig:AMRtest_noTI}
     \end{subfigure}
     \hfill
     \begin{subfigure}[b]{0.35\textwidth}
         \centering
         \includegraphics[width=0.95\textwidth]{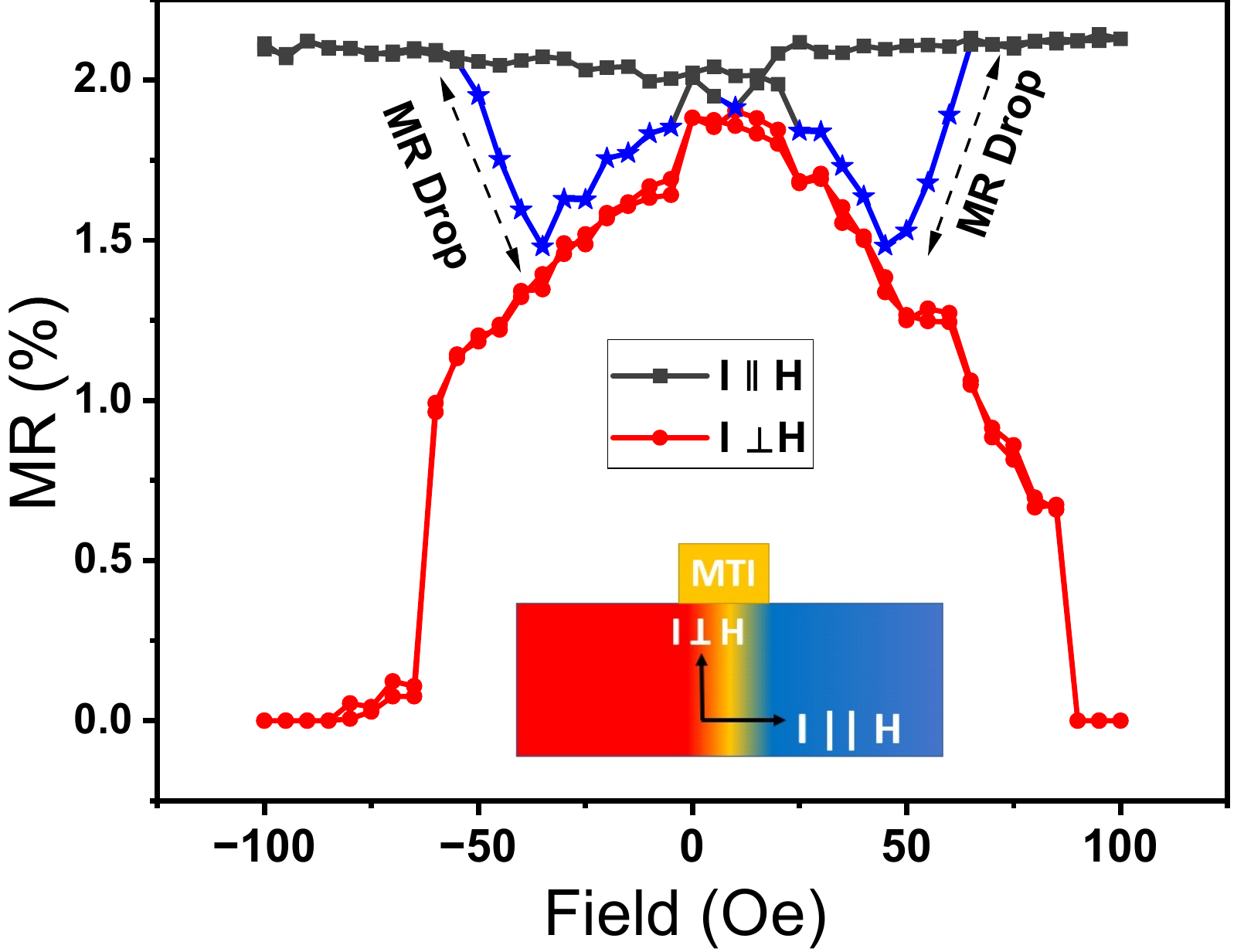}
         \caption{AMR with TI}
         \label{fig:AMRtest_TI}
     \end{subfigure}
        \caption{Demostration of domain wall pinning at the FM-MTI interface due to direct exchange interaction.}
        \label{fig:AMRtet}
\end{figure}

\noindent To realize domain wall pinning at the crossbar locations, we measured the anisotropic magneto-resistance (AMR), which is determined by the angle between the flow of conduction electrons and the local magnetic moments. Initially, we calculated the AMR of a standalone \foo nife nanowire with a domain wall introduced at the center (see inset of Fig.~\ref{fig:AMRtest_noTI}). An external current of 3x10\textsuperscript{7} A/cm\textsuperscript{2} flows in the +x-direction, while a magnetic field ranging from –100 to 100 Oe with a step size of 5 Oe is applied parallel and perpendicular to the current flow (+x and +y-directions, respectively). For each value of applied field, the simulation time is 1 ns allowing enough time for energy minimum magnetization state. Fig.\ref{fig:AMRtet} shows how the magnetoresistance (MR) changes under these conditions.

When the magnetic field is applied parallel to the current, the MR of the nanowire (NW) remains high due to increased s-d scattering of electrons. The black lines in Fig.~\ref{fig:AMRtet} show the magnetoresistance of the nanowire when the magnetic field is applied parallel to the direction of current flow.  Since the easy axis of the nanowire is along the +x-direction, the MR remains nearly constant with increasing field. In contrast, a magnetic field applied perpendicular to the current results in a steady drop in MR. The red line shows the MR for this scenario. After the coercive field is reached, the MR plateaus, indicating that the magnetization of the nanowire is fully aligned along the +y-direction. These results are consistent with a similar experiment reported in~\cite{jin2017tuning}.

Next, we considered the NW with the magnetic TI layer on top. A similar drop in resistance is observed when the field is applied perpendicular to the current. However, a different behavior emerges when the magnetic field is applied along the current direction. In this case, we observed two dips in magnetoresistance at approximately 60~Oe and 50~Oe. These dips correspond to domain wall pinning at the crosspoints, indicating that a stronger field is required to align the NW magnetization along the +x-direction.

\subsection{Depining of Domain Walls}
\label{sec:res_depin}

\begin{figure}[H]
     \centering
     \begin{subfigure}[b]{0.24\textwidth}
         \centering
        \includegraphics[width=0.98\textwidth]{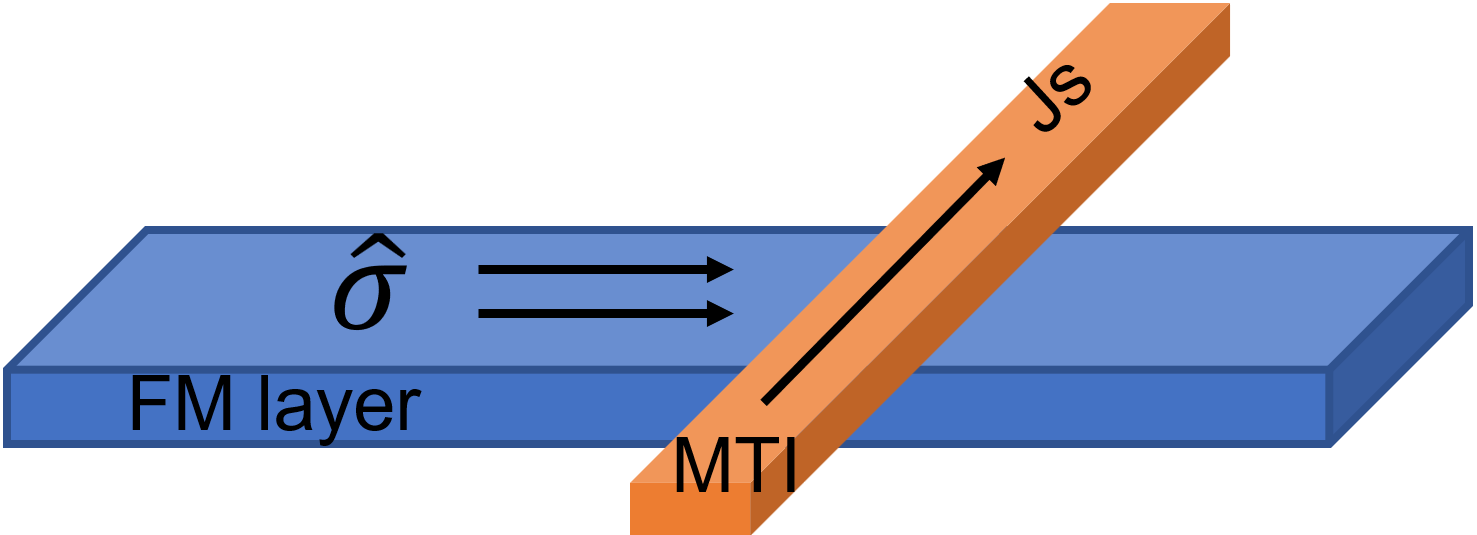}
         \caption{MTI-FM intersection}
         \label{fig:depin1}
     \end{subfigure}
     \hfill
     \begin{subfigure}[b]{0.24\textwidth}
         \centering
         \includegraphics[width=0.98\textwidth]{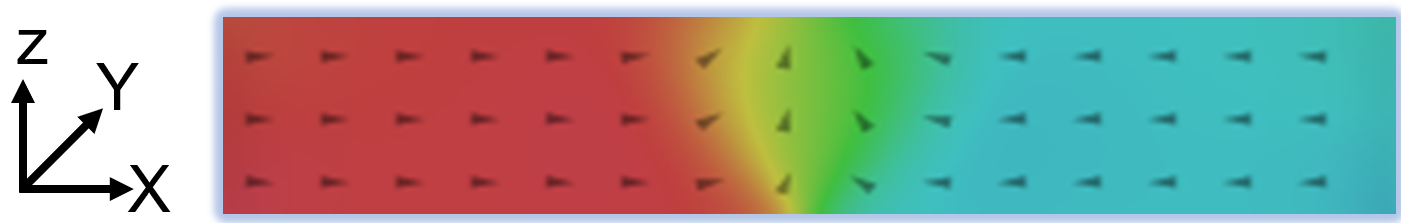}
         \caption{HTH DW in FM NW}
         \label{fig:depin2}
     \end{subfigure}
        \caption{Domain wall depinning at the MTI-FM interface.}
        \label{fig:depining}
\end{figure}

\noindent The Slonczewski-like torque inside the FM layer, due to spin diffusion from the MTI surface, can depin domain walls pinned at the intersection of the MTI layer and the FM nanowire. Fig.~\ref{fig:depin1} shows the schematic of the intersection geometry of the proposed device, and Fig.~\ref{fig:depin2} shows the head-to-head (HTH) domain wall magnetization in the nanowire. The effective field ($\vec{H}_{\mathrm{STT}}$) associated with the torque, derived in (\ref{eqn:T_STT}), can be expressed as
\begin{equation}
\vec{H}_{\mathrm{STT}} = T_{\mathrm{STT}}(\vec{m} \times \hat{\sigma}),
\label{eqn:H_STT}
\end{equation}
where $\hat{\sigma} = \hat{j} \times \hat{k}$, and $\hat{j}$ and $\hat{z}$ are unit vectors in the current and out-of-plane directions, respectively. The polarization vector in this device structure lies in the x-direction, and the effective field aligns with it. Therefore, the field is parallel to one of the domains in the nanowire, indicating that the torque can influence the domain walls. For PMA nanowires with Néel walls, this x-directed field has a similar effect on the walls; thus, the depinning method remains identical.

In this structure, the $T_{STT}$ torque acts on the domain walls only at the intersection; therefore, there is no steady-state domain wall motion due to this torque alone. We consider that the spin efficiency at the interface, $p = 1$. In Permalloy, considering the spin diffusion length, $\lambda_{sf} = 5$ nm~\cite{mellnik2014spin}, spin decoherence length, $\lambda_\phi = 1$ nm, and spin precession length, $\lambda_j = 1$ nm, we have calculated the in-plane and out-of-plane torque to be $5.9\times10^{-4}$ T and $1.2\times10^{-4}$ T, respectively, from Eq.~\ref{eqn:T_STT}. We incorporated this torque into the micromagnetic simulation as an external field ($\vec{H}_{STT}$) at the intersection to mimic the spin transfer. 

Without any external excitations in the MTI nanobars, the intersections create maximum pinning potential. During current flow through the MTI nanobars, the pinning strength is reduced due to the inserted spin torque into the nanowire. By controlling the current through the MTI nanobars, variable pinning strength can be achieved, which is of great importance in numerous computing applications. First, we calculated the critical current requirement based on Eq.~\ref{eq:LLG}, assuming that thermal fluctuations are negligible. In this case, there is no current flow through the nanobars. While our experiments varied the width of the nanobars from 20~nm to 80~nm in 10~nm steps, the nanowire's width, thickness, and length were kept constant.

\begin{figure}[H]
\centering
\includegraphics[width=2.8in]{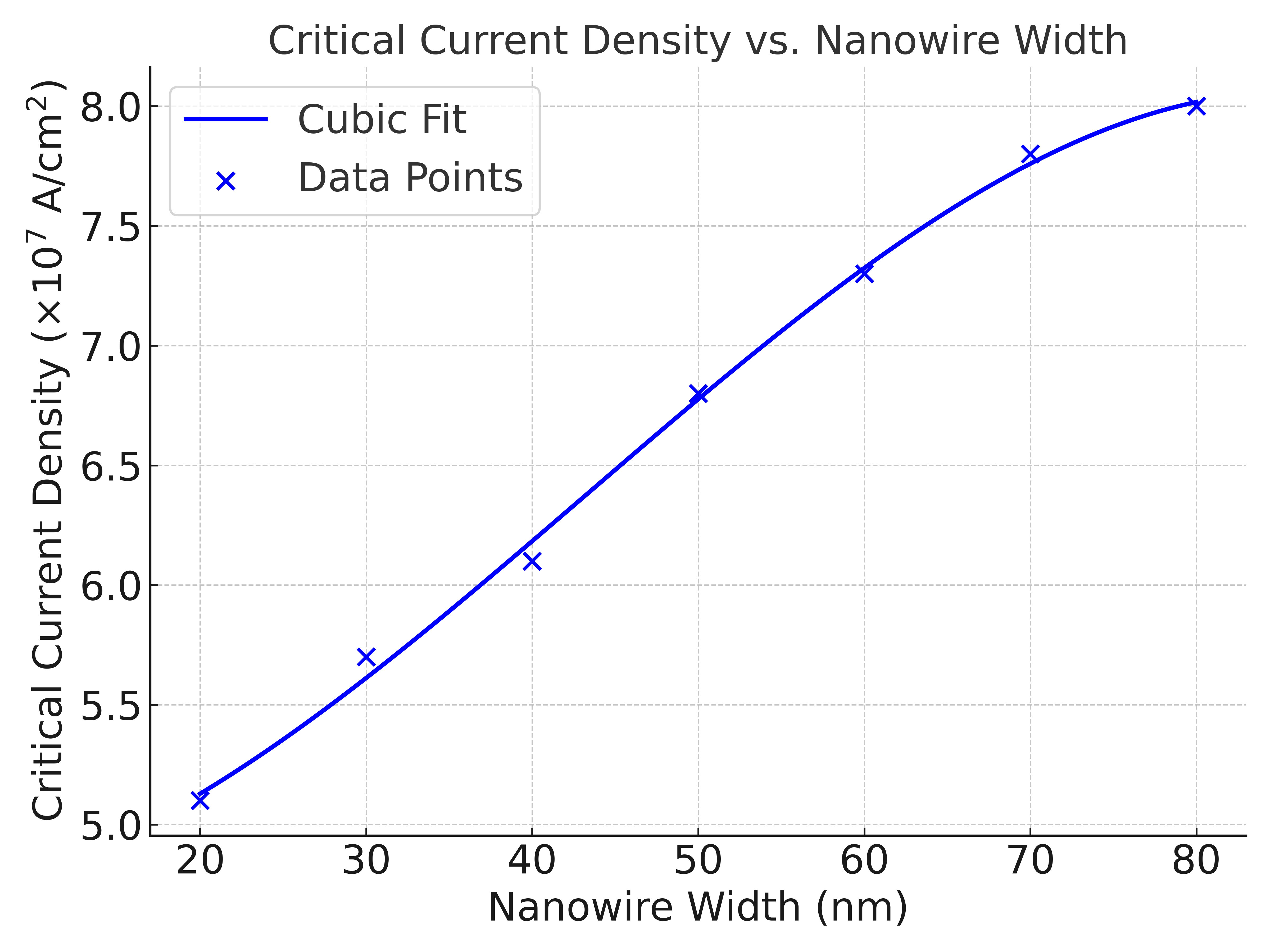}
\caption{Critical shift current increases with the width of MTI.}
\label{fig:width_vs_current}
\end{figure}

\noindent We initialized the nanowire with a domain wall at the center and allowed the nanowire to relax. To calibrate DW movement, we applied current pulses of 1 ns and tracked the DW position along the nanowire after relaxation. Fig.~\ref{fig:width_vs_current} shows how the critical current requirement for shifting varies with the width of the MTI nanobars without any current flowing through them. With increased width of MTI nanobars, the current required for shifting increases almost linearly, which implies that the pinning potential increases accordingly. This is expected, as the intersection area increases with the width of MTI nanobars, which in turn increases the exchange interaction between FM and MTI, causing the increased pinning potential. We tuned the shift current to find the critical current density for shifting in this ideal scenario for each width of the MTI nanobars. A nanowire containing a 20~nm wide nanobar requires a minimum current density of 5.1×10\textsuperscript{7}~A/cm\textsuperscript{2} to successfully depin the wall. We also calculated the shift current density for an 80~nm wide nanobar as 7.9×10\textsuperscript{7}~A/cm\textsuperscript{2}. These values serve as the control for the proposed device.

\subsection{Performance of the FM-MTI DWM}

\noindent Pinning domain walls using magnetic topological insulator layers enables controllable pinning and localized spin torque-assisted depinning. This makes the FM-MTI nanowire a promising candidate for reconfigurable memory devices with low-power operation compared to state-of-the-art IMA domain wall memories and recently proposed devices with alternative pinning methods. Since the performance depends on the nanowires’ materials, shape, and dimensions, direct comparison between them is often inconclusive. Also, the critical current for shifting domain walls depends on the current pulse width and the number of domains in the nanowire. Table~\ref{tab:comparison} compares the performance of FM-MTI nanobars with typical IMA domain wall memories and recently proposed DWMs with alternative pinning methods.

\begin{table}[H]
\centering
\caption{Comparison Between Domain Wall Memories.}
\begin{tabular}{|l|l|l|l|l|}
\hline
Metrics & \begin{tabular}[c]{@{}l@{}}IMA \\ DWM~\cite{al2023multi}\end{tabular} & \begin{tabular}[c]{@{}l@{}}Ga+ ion \\ ~\cite{giuliano2023ga+}\end{tabular} & \begin{tabular}[c]{@{}l@{}}DI-based \\ DWM~\cite{lee2023position}\end{tabular} & \begin{tabular}[c]{@{}l@{}}FM-MTI\\ (this work)\end{tabular} \\ \hline
\begin{tabular}[c]{@{}l@{}}Tunable \\ Pinning\end{tabular} & No & Yes & Yes & Yes \\ \hline
\begin{tabular}[c]{@{}l@{}}Pinning \\ Method\end{tabular} & \begin{tabular}[c]{@{}l@{}}Shape\\ Deformity\end{tabular} & \begin{tabular}[c]{@{}l@{}}Material\\ Properties\end{tabular} & \begin{tabular}[c]{@{}l@{}}Dipolar \\ Interaction\end{tabular} & \begin{tabular}[c]{@{}l@{}}Exchange\\ Interaction\end{tabular} \\ \hline
Wall Type & Vortex & Néel & Néel & Transverse \\ \hline
\begin{tabular}[c]{@{}l@{}}Shift\\ Current (A/m\textsuperscript{2})\end{tabular} & $7.7 \times 10^{11}$ & $8.5 \times 10^{11}$ & $3 \times 10^{11}$ & $5.1 \times 10^{11}$ \\ \hline
DW Speed & 250 m/s & NM & 100 m/s & $\sim$275 m/s \\ \hline
\end{tabular}
\label{tab:comparison}
\end{table}

\noindent Recently, Bahri~\cite{al2023multi} calculated the critical parameters of domain wall memory with in-plane magnetic anisotropy (IMA) materials, where segmentation of the nanowire creates the pinning potential. The proposed device in this work demonstrates better speed—34\% higher than conventional IMA nanowires—while requiring 15\% lower shift current. Moreover, the area of the FM–MTI nanowire is 40$\times$ smaller than that reported in~\cite{al2023multi}. These key metrics indicate that the proposed device can serve as a low-power alternative to conventional domain wall memory devices, with a controllable pinning mechanism. The combination of electrically tunable pinning strength and low-power operation offered by the FM–MTI domain wall memory can benefit data-intensive non-Boolean computation.

\begin{figure}[!t]
\centering
\includegraphics[width=3in]{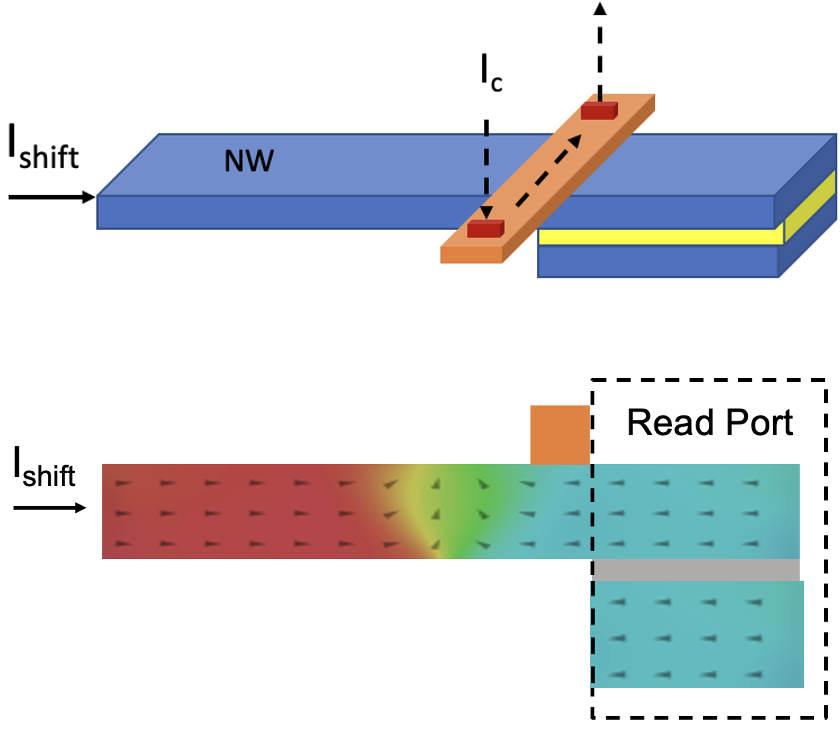}
\caption{Sigmoid function realization by controllable domain wall motion.}
\label{fig:sigmoid}
\end{figure}

\begin{figure}[!b]
\centering
\includegraphics[width=2.8in]{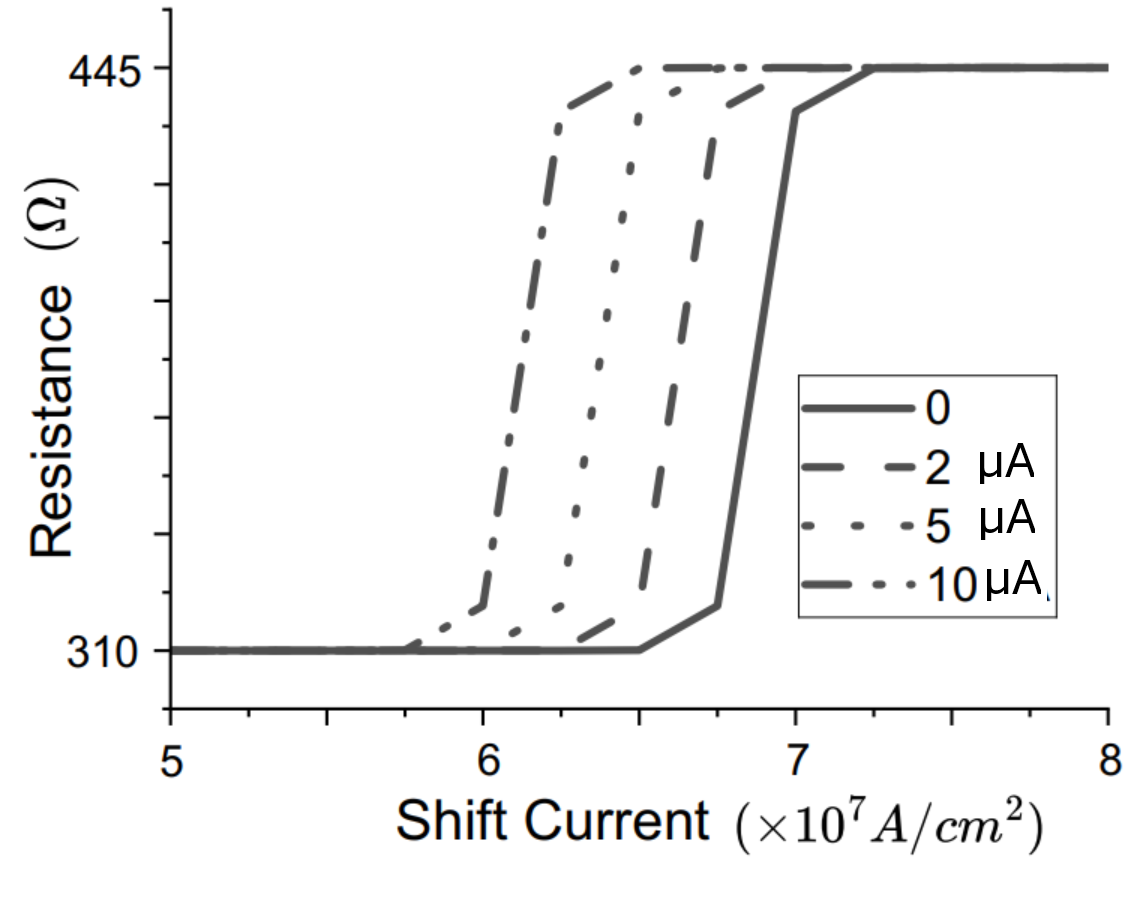}
\caption{Controllable pinning strength can generate sigmoid response in FM-MTI domain wall memory.}
\label{fig:control}
\end{figure}

\section{Activation Function by an FM-MTI Device}

In a simplified architecture of an artificial neural network, an activation function (AF) operates on the weighted sum of inputs to a neuron and determines the output ($Y$) of the neuron as

\begin{equation}
    Y = f_a\left(\sum w_i x_i + b\right)
    \label{eq:ANN}
\end{equation}

\noindent where $w_i$ is the weight corresponding to the $i^{\text{th}}$ input $x_i$; the bias $b$ denotes a weight connected to an input of fixed value 1; and $f_a$ is the activation function. While software implementation of such neural networks is straightforward, it often requires complex, power-hungry hardware, and the runtime typically increases non-linearly with input size. On the other hand, hardware implementation of neurons has been encouraged due to better runtime—for example, in~\cite{bhanja2016non}, the runtime does not increase with input size.

The threshold (or hard-limiting) function is popular due to its ease of implementation in CMOS or spintronic hardware. However, various non-linear functions such as $\tanh$ and ReLU are also widely used. The thresholding activation function determines the output of the neuron as:

\[
    Y = 
\begin{cases}
    1,& \text{if } \sum w_i x_i + b \geq 0 \\
    0,              & \text{otherwise}
\end{cases}
\]

\noindent The FM-MTI domain wall nanowire can realize this activation function using controllable pinning strength. To implement the function, an FM nanowire is considered with only one MTI nanobar as a pinning site (Fig.~\ref{fig:sigmoid}). A typical read port is placed to the right of the pinning location. The weighted sum of inputs in the form of electrical current is passed through the nanobars, and a shift current less than the critical current flows through the FM nanowire.

The nanowire is initialized with a magnetization such that the read port shows low resistance due to the parallel orientation of the fixed layer and the domain wall magnetization above it. Since the shift current is below the critical threshold, the domain wall does not pass beyond the pinning site, and there is no change in resistance. If the weighted sum of inputs exceeds the threshold, the current through the MTI layer reduces the pinning strength. Consequently, the wall moves, and the magnetization at the read port switches. The resistance increases due to the mismatch between the magnetizations.

\noindent Fig.~\ref{fig:control} shows the resistance profile of the device in Fig.~\ref{fig:sigmoid}. The device size is similar to that in Fig.~\ref{fig:Tilt}, and resistance is calculated using the method described in~\cite{roxy2018reading}. Fig.~\ref{fig:control} demonstrates how the current through the MTI nanobar controls the resistance at the read port. Initially, the free layer (in the nanowire) has parallel magnetization to the fixed layer, resulting in a low resistance of 310~$\Omega$. In the absence of current through the MTI nanowire, the pinning potential is at its maximum, as no torque is applied from the MTI to depin the wall. In this case, the shift current required to move the wall is the highest (solid line in Fig.~\ref{fig:control}), around 7.1×10\textsuperscript{7}~A/cm\textsuperscript{2}, consistent with the value shown in Fig.~\ref{fig:width_vs_current}. The surface charge at both ends of the nanowire is removed, similar to the previous experiment, to mimic an infinitely long nanowire. When the wall moves approximately 250~nm, the fixed layer becomes +x-directed, resulting in a higher resistance of 445~$\Omega$ at the read port due to the anti-parallel configuration between the free and fixed layers.

When current flows through the MTI nanobar, the $T_{\mathrm{STT}}$ torque from the MTI layer acts on the nanowire, forcing the wall to move from the pinned location. As a result, the pinning potential is reduced, and a lower shift current is sufficient to move the domain wall. For the MTI currents shown in the inset of Fig.~\ref{fig:control}, the corresponding shift current requirements (dotted lines) are all below 7.1×10\textsuperscript{7}~A/cm\textsuperscript{2}. This behavior produces a resistance profile at the read port that mimics a threshold activation function, where the resistance values represent the neuron's output.

In this work, we demonstrated the functionality of a domain wall memory with electrically reconfigurable pinning strength. This enables better control of pinning potentials and locations, which is useful in non-Boolean computations. Future work will focus on achieving finer granularity of control for near-analog computation.

\bibliographystyle{IEEEtran}
\bibliography{TI,DW, NEW}

\begin{thebibliography}{10}
\providecommand{\url}[1]{#1}
\csname url@samestyle\endcsname
\providecommand{\newblock}{\relax}
\providecommand{\bibinfo}[2]{#2}
\providecommand{\BIBentrySTDinterwordspacing}{\spaceskip=0pt\relax}
\providecommand{\BIBentryALTinterwordstretchfactor}{4}
\providecommand{\BIBentryALTinterwordspacing}{\spaceskip=\fontdimen2\font plus
\BIBentryALTinterwordstretchfactor\fontdimen3\font minus
  \fontdimen4\font\relax}
\providecommand{\BIBforeignlanguage}[2]{{%
\expandafter\ifx\csname l@#1\endcsname\relax
\typeout{** WARNING: IEEEtran.bst: No hyphenation pattern has been}%
\typeout{** loaded for the language `#1'. Using the pattern for}%
\typeout{** the default language instead.}%
\else
\language=\csname l@#1\endcsname
\fi
#2}}
\providecommand{\BIBdecl}{\relax}
\BIBdecl

\bibitem{parkin2008magnetic}
S.~S. Parkin, M.~Hayashi, and L.~Thomas, ``Magnetic domain-wall racetrack
  memory,'' \emph{Science}, vol. 320, no. 5873, pp. 190--194, 2008.

\bibitem{akerman2010stochastic}
J.~Akerman, M.~Mu{\~n}oz, M.~Maicas, and J.~L. Prieto, ``Stochastic nature of
  the domain wall depinning in permalloy magnetic nanowires,'' \emph{Physical
  Review B}, vol.~82, no.~6, p. 064426, 2010.

\bibitem{huang2009domain}
S.-H. Huang and C.-H. Lai, ``Domain-wall depinning by controlling its
  configuration at notch,'' \emph{Applied Physics Letters}, vol.~95, no.~3,
  2009.

\bibitem{zhou2017current}
X.~Zhou, Z.~Huang, W.~Zhang, Y.~Yin, P.~D{\"u}rrenfeld, S.~Dong, and Y.~Zhai,
  ``Current-induced multiple domain wall motion modulated by magnetic pinning
  in zigzag shaped nanowires,'' \emph{AIP Advances}, vol.~7, no.~5, 2017.

\bibitem{roxy2022pinning}
K.~Roxy, S.~Longofono, S.~Olliver, S.~Bhanja, and A.~K. Jones, ``Pinning fault
  mode modeling for dwm shifting,'' \emph{IEEE Transactions on Circuits and
  Systems II: Express Briefs}, vol.~69, no.~7, pp. 3319--3323, 2022.

\bibitem{polenciuc2014domain}
I.~Polenciuc, A.~J. Vick, D.~Allwood, T.~Hayward, G.~Vallejo-Fernandez,
  K.~O'Grady, and A.~Hirohata, ``Domain wall pinning for racetrack memory using
  exchange bias,'' \emph{Applied Physics Letters}, vol. 105, no.~16, p. 162406,
  2014.

\bibitem{king2014local}
J.~King, A.~Ganguly, D.~Burn, S.~Pal, E.~Sallabank, T.~Hase, A.~Hindmarch,
  A.~Barman, and D.~Atkinson, ``Local control of magnetic damping in
  ferromagnetic/non-magnetic bilayers by interfacial intermixing induced by
  focused ion-beam irradiation,'' \emph{Applied physics letters}, vol. 104,
  no.~24, 2014.

\bibitem{burn2014control}
D.~Burn and D.~Atkinson, ``Control of domain wall pinning by localised focused
  ga+ ion irradiation on au capped nife nanowires,'' \emph{Journal of Applied
  Physics}, vol. 116, no.~16, p. 163901, 2014.

\bibitem{rantschler2007effect}
J.~Rantschler, R.~McMichael, A.~Castillo, A.~Shapiro, W.~Egelhoff,
  B.~Maranville, D.~Pulugurtha, A.~Chen, and L.~Connors, ``Effect of 3d, 4d,
  and 5d transition metal doping on damping in permalloy thin films,''
  \emph{Journal of applied physics}, vol. 101, no.~3, 2007.

\bibitem{faulkner2007rapid}
C.~Faulkner, D.~Atkinson, D.~Allwood, and R.~Cowburn, ``Rapid tuning of
  ni81fe19/au bilayer magnetic properties by focused ion beam intermixing,''
  \emph{Journal of magnetism and magnetic materials}, vol. 319, no. 1-2, pp.
  9--12, 2007.

\bibitem{jin2017tuning}
T.~Jin, M.~Ranjbar, S.~He, W.~C. Law, T.~Zhou, W.~Lew, X.~Liu, and
  S.~Piramanayagam, ``Tuning magnetic properties for domain wall pinning via
  localized metal diffusion,'' \emph{Scientific reports}, vol.~7, no.~1, p.
  16208, 2017.

\bibitem{breitkreutz2014controlled}
S.~Breitkreutz, I.~Eichwald, J.~Kiermaier, G.~Hiblot, G.~Csaba, W.~Porod,
  D.~Schmitt-Landsiedel, and M.~Becherer, ``Controlled domain wall pinning in
  nanowires with perpendicular magnetic anisotropy by localized fringing
  fields,'' \emph{Journal of Applied Physics}, vol. 115, no.~17, 2014.

\bibitem{franken2013voltage}
J.~Franken, Y.~Yin, A.~Schellekens, A.~van~den Brink, H.~Swagten, and
  B.~Koopmans, ``Voltage-gated pinning in a magnetic domain-wall conduit,''
  \emph{Applied Physics Letters}, vol. 103, no.~10, 2013.

\bibitem{bernand2013electric}
A.~Bernand-Mantel, L.~Herrera-Diez, L.~Ranno, S.~Pizzini, J.~Vogel, D.~Givord,
  S.~Auffret, O.~Boulle, I.~M. Miron, and G.~Gaudin, ``Electric-field control
  of domain wall nucleation and pinning in a metallic ferromagnet,''
  \emph{Applied Physics Letters}, vol. 102, no.~12, 2013.

\bibitem{metaxas2009periodic}
P.~Metaxas, P.-J. Zermatten, J.-P. Jamet, J.~Ferr{\'e}, G.~Gaudin, B.~Rodmacq,
  A.~Schuhl, and R.~Stamps, ``Periodic magnetic domain wall pinning in an
  ultrathin film with perpendicular anisotropy generated by the stray magnetic
  field of a ferromagnetic nanodot array,'' \emph{Applied Physics Letters},
  vol.~94, no.~13, 2009.

\bibitem{o2011tunable}
L.~O’Brien, D.~Petit, E.~Lewis, R.~Cowburn, D.~Read, J.~Sampaio, H.~Zeng, and
  A.-V. Jausovec, ``Tunable remote pinning of domain walls in magnetic
  nanowires,'' \emph{Physical Review Letters}, vol. 106, no.~8, p. 087204,
  2011.

\bibitem{hiramatsu2013domain}
R.~Hiramatsu, T.~Koyama, H.~Hata, T.~Ono, D.~Chiba, S.~Fukami, and N.~Ishiwata,
  ``Domain wall pinning by a stray field from nife on a co/ni nanowire,''
  \emph{Journal of the Korean Physical Society}, vol.~63, pp. 608--611, 2013.

\bibitem{metaxas2013spatially}
P.~Metaxas, P.-J. Zermatten, R.~Novak, S.~Rohart, J.-P. Jamet, R.~Weil,
  J.~Ferr{\'e}, A.~Mougin, R.~Stamps, G.~Gaudin \emph{et~al.}, ``Spatially
  periodic domain wall pinning potentials: Asymmetric pinning and dipolar
  biasing,'' \emph{Journal of Applied Physics}, vol. 113, no.~7, 2013.

\bibitem{franken2014beam}
J.~H. Franken, M.~A. van~der Heijden, T.~H. Ellis, R.~Lavrijsen, C.~Daniels,
  D.~McGrouther, H.~J. Swagten, and B.~Koopmans, ``Beam-induced fe nanopillars
  as tunable domain-wall pinning sites,'' \emph{Advanced Functional Materials},
  vol.~24, no.~23, pp. 3508--3514, 2014.

\bibitem{van2014control}
R.~Van~Mourik, C.~Rettner, B.~Koopmans, and S.~Parkin, ``Control of domain wall
  pinning by switchable nanomagnet state,'' \emph{Journal of Applied Physics},
  vol. 115, no.~17, 2014.

\bibitem{hurst2017reconfigurable}
A.~C. Hurst, J.~A. Izaac, F.~Altaf, V.~Baltz, and P.~J. Metaxas,
  ``Reconfigurable magnetic domain wall pinning using vortex-generated magnetic
  fields,'' \emph{Applied Physics Letters}, vol. 110, no.~18, 2017.

\bibitem{lee2023position}
T.~Lee, S.~Jeong, S.~Kim, and K.-J. Kim, ``Position-reconfigurable pinning for
  magnetic domain wall motion,'' \emph{Scientific Reports}, vol.~13, no.~1, p.
  6791, 2023.

\bibitem{liu2010model}
C.-X. Liu, X.-L. Qi, H.~Zhang, X.~Dai, Z.~Fang, and S.-C. Zhang, ``Model
  hamiltonian for topological insulators,'' \emph{Physical Review B}, vol.~82,
  no.~4, p. 045122, 2010.

\bibitem{zhang2009topological}
H.~Zhang, C.-X. Liu, X.-L. Qi, X.~Dai, Z.~Fang, and S.-C. Zhang, ``Topological
  insulators in bi2se3, bi2te3 and sb2te3 with a single dirac cone on the
  surface,'' \emph{Nature physics}, vol.~5, no.~6, pp. 438--442, 2009.

\bibitem{mellnik2014spin}
A.~Mellnik, J.~Lee, A.~Richardella, J.~Grab, P.~Mintun, M.~H. Fischer,
  A.~Vaezi, A.~Manchon, E.-A. Kim, N.~Samarth \emph{et~al.}, ``Spin-transfer
  torque generated by a topological insulator,'' \emph{Nature}, vol. 511, no.
  7510, pp. 449--451, 2014.

\bibitem{edelstein1990spin}
V.~M. Edelstein, ``Spin polarization of conduction electrons induced by
  electric current in two-dimensional asymmetric electron systems,''
  \emph{Solid State Communications}, vol.~73, no.~3, pp. 233--235, 1990.

\bibitem{fan2014magnetization}
Y.~Fan, P.~Upadhyaya, X.~Kou, M.~Lang, S.~Takei, Z.~Wang, J.~Tang, L.~He, L.-T.
  Chang, M.~Montazeri \emph{et~al.}, ``Magnetization switching through giant
  spin--orbit torque in a magnetically doped topological insulator
  heterostructure,'' \emph{Nature materials}, vol.~13, no.~7, pp. 699--704,
  2014.

\bibitem{fischer2016spin}
M.~H. Fischer, A.~Vaezi, A.~Manchon, and E.-A. Kim, ``Spin-torque generation in
  topological insulator based heterostructures,'' \emph{Physical Review B},
  vol.~93, no.~12, p. 125303, 2016.

\bibitem{yokoyama2010theoretical}
T.~Yokoyama, J.~Zang, and N.~Nagaosa, ``Theoretical study of the dynamics of
  magnetization on the topological surface,'' \emph{Physical Review B},
  vol.~81, no.~24, p. 241410, 2010.

\bibitem{manchon2012spin}
A.~Manchon, R.~Matsumoto, H.~Jaffres, and J.~Grollier, ``Spin transfer torque
  with spin diffusion in magnetic tunnel junctions,'' \emph{Physical Review B},
  vol.~86, no.~6, p. 060404, 2012.

\bibitem{kurebayashi2014antidamping}
H.~Kurebayashi, J.~Sinova, D.~Fang, A.~Irvine, T.~Skinner, J.~Wunderlich,
  V.~Nov{\'a}k, R.~Campion, B.~Gallagher, E.~Vehstedt \emph{et~al.}, ``An
  antidamping spin--orbit torque originating from the berry curvature,''
  \emph{Nature nanotechnology}, vol.~9, no.~3, pp. 211--217, 2014.

\bibitem{keller2012facing}
R.~Keller, D.~Kramer, and J.-P. Weiss, \emph{Facing the Multicore-Challenge II:
  Aspects of New Paradigms and Technologies in Parallel Computing}.\hskip 1em
  plus 0.5em minus 0.4em\relax Springer, 2012, vol. 7174.

\bibitem{zhao2012magnetic}
W.~Zhao, Y.~Zhang, H.~Trinh, J.~Klein, C.~Chappert, R.~Mantovan, A.~Lamperti,
  R.~Cowburn, T.~Trypiniotis, M.~Klaui \emph{et~al.}, ``Magnetic domain-wall
  racetrack memory for high density and fast data storage,'' in
  \emph{Solid-State and Integrated Circuit Technology (ICSICT), 2012 IEEE 11th
  International Conference on}.\hskip 1em plus 0.5em minus 0.4em\relax IEEE,
  2012, pp. 1--4.

\bibitem{foerster2014domain}
M.~Foerster, O.~Boulle, S.~Esefelder, R.~Mattheis, and M.~Kl{\"a}ui, ``Domain
  wall memory device,'' \emph{Handbook of Spintronics}, pp. 1--46.

\bibitem{venkatesan2013dwm}
R.~Venkatesan, M.~Sharad, K.~Roy, and A.~Raghunathan, ``Dwm-tapestri-an energy
  efficient all-spin cache using domain wall shift based writes,'' in
  \emph{2013 Design, Automation \& Test in Europe Conference \& Exhibition
  (DATE)}.\hskip 1em plus 0.5em minus 0.4em\relax IEEE, 2013, pp. 1825--1830.

\bibitem{2013sunDac}
Z.~Sun, W.~Wu, and H.~Li, ``Cross-layer racetrack memory design for ultra high
  density and low power consumption,'' in \emph{Design Automation Conference
  (DAC), 2013 50th ACM/EDAC/IEEE}, pp. 1--6.

\bibitem{berger1984exchange}
L.~Berger, ``Exchange interaction between ferromagnetic domain wall and
  electric current in very thin metallic films,'' \emph{Journal of Applied
  Physics}, vol.~55, no.~6, pp. 1954--1956, 1984.

\bibitem{tatara2004theory}
G.~Tatara and H.~Kohno, ``Theory of current-driven domain wall motion: Spin
  transfer versus momentum transfer,'' \emph{Physical review letters}, vol.~92,
  no.~8, p. 086601, 2004.

\bibitem{tatara2008microscopic}
G.~Tatara, H.~Kohno, and J.~Shibata, ``Microscopic approach to current-driven
  domain wall dynamics,'' \emph{Physics Reports}, vol. 468, no.~6, pp.
  213--301, 2008.

\bibitem{kumar2018domain}
D.~Kumar, T.~Jin, S.~Al~Risi, R.~Sbiaa, W.~S. Lew, and S.~Piramanayagam,
  ``Domain wall motion control for racetrack memory applications,'' \emph{IEEE
  Transactions on Magnetics}, vol.~55, no.~3, pp. 1--8, 2018.

\bibitem{mumax}
A.~Vansteenkiste, J.~Leliaert, M.~Dvornik, M.~Helsen, F.~Garcia-Sanchez, and
  B.~Van~Waeyenberge, ``The design and verification of mumax3,'' \emph{AIP
  advances}, vol.~4, no.~10, p. 107133, 2014.

\bibitem{reza2019modeling}
A.~K. Reza, ``Modeling and simulation of topological insulators, topological
  semi-metals and ferrimagnets for time and energy efficient switching of
  magnetic tunnel junction,'' Ph.D. dissertation, Purdue University, 2019.

\bibitem{zhang2014electrical}
M.~Zhang, L.~Lv, Z.~Wei, L.~Yang, X.~Yang, and Y.~Zhao, ``Electrical and
  magnetic transport properties of co-doped bi 2 se 3 topological insulator
  crystals,'' \emph{International Journal of Modern Physics B}, vol.~28,
  no.~17, p. 1450108, 2014.

\bibitem{kumar2025co}
R.~Kumar, D.~Sisodiya, K.~Vijay, S.~Banik, S.~Sen, P.~Babu, and
  D.~Bhattacharyya, ``Co doped bi2se3 topological insulator: a combined
  theoretical and experimental study,'' \emph{Materials Science in
  Semiconductor Processing}, vol. 195, p. 109572, 2025.

\bibitem{liu2023magnetic}
J.~Liu and T.~Hesjedal, ``Magnetic topological insulator heterostructures: A
  review,'' \emph{Advanced Materials}, vol.~35, no.~27, p. 2102427, 2023.

\bibitem{abo2013definition}
G.~S. Abo, Y.-K. Hong, J.~Park, J.~Lee, W.~Lee, and B.-C. Choi, ``Definition of
  magnetic exchange length,'' \emph{IEEE Transactions on Magnetics}, vol.~49,
  no.~8, pp. 4937--4939, 2013.

\bibitem{yan2024rules}
Q.~Yan, H.~Li, H.~Jiang, Q.-F. Sun, and X.~Xie, ``Rules for dissipationless
  topotronics,'' \emph{Science Advances}, vol.~10, no.~23, p. eado4756, 2024.

\bibitem{jin2019tilted}
T.~Jin, F.~Tan, C.~C.~I. Ang, W.~Gan, J.~Cao, W.~S. Lew, and S.~Piramanayagam,
  ``Tilted magnetisation for domain wall pinning in racetrack memory,''
  \emph{Journal of Magnetism and Magnetic Materials}, vol. 489, p. 165410,
  2019.

\bibitem{al2023multi}
M.~Al~Bahri, M.~Al~Hinaai, and T.~Al~Harthy, ``Multi-segmented nanowires for
  vortex magnetic domain wall racetrack memory,'' \emph{Chinese Physics B},
  vol.~32, no.~12, p. 127508, 2023.

\bibitem{giuliano2023ga+}
D.~Giuliano, L.~Gnoli, V.~Ahrens, M.~R. Roch, M.~Becherer, G.~Turvani,
  M.~Vacca, and F.~Riente, ``Ga+ ion irradiation-induced tuning of artificial
  pinning sites to control domain wall motion,'' \emph{ACS Applied Electronic
  Materials}, vol.~5, no.~2, pp. 985--993, 2023.

\bibitem{bhanja2016non}
S.~Bhanja, D.~Karunaratne, R.~Panchumarthy, S.~Rajaram, and S.~Sarkar,
  ``Non-boolean computing with nanomagnets for computer vision applications,''
  \emph{Nature nanotechnology}, vol.~11, no.~2, pp. 177--183, 2016.

\bibitem{roxy2018reading}
K.~A. Roxy and S.~Bhanja, ``Reading nanomagnetic energy minimizing
  coprocessor,'' \emph{IEEE Transactions on Nanotechnology}, vol.~17, no.~2,
  pp. 368--372, 2018.

\end{thebibliography}
\end{document}